\documentclass{aa}

\usepackage[varg]{txfonts}

\usepackage{lineno}
\usepackage{natbib}
\usepackage{amsmath}
\usepackage{graphicx}
\usepackage{txfonts}
\usepackage{xspace}
\usepackage{xcolor}
\usepackage{textcomp}
\usepackage{multirow}
\usepackage{hyperref}
\hypersetup{
    colorlinks=true,
    linkcolor={red!50!black},
    citecolor={blue!70!black},
    urlcolor={blue!80!black}
}

\newcommand{\gro}{\object{GRO\,J1744$-$28}\xspace}

\newcommand{\nustar}{\mbox{\textsl{NuSTAR}}\xspace}
\newcommand{\swift}{\textsl{Swift}\xspace}
\newcommand{\integral}{\textsl{INTEGRAL}\xspace}
\newcommand{\chandra}{\textsl{Chandra}\xspace}
\newcommand{\xmm}{\textsl{XMM-Newton}\xspace}
\newcommand{\sax}{\textsl{BeppoSAX}\xspace}
\newcommand{\rxte}{\textsl{RXTE}\xspace}
\newcommand{\nicer}{\textsl{NICER}\xspace}

\newcommand{\fdcut}{\texttt{FDcut}\xspace}
\newcommand{\cutoffpl}{\texttt{cutoffpl}\xspace}
\newcommand{\npex}{\texttt{NPEX}\xspace}
\newcommand{\gabs}{\texttt{gabs}\xspace}

\newcommand{\asec}{\ensuremath{''}\xspace}

\newcommand{\snr}{S/N\xspace}
\newcommand{\kev}{\ensuremath{\text{keV}}\xspace}
\newcommand{\msun}{\ensuremath{\text{M}_{\odot}}\xspace}

\newcommand{\feka}{\ensuremath{\mathrm{Fe}~\mathrm{K}\alpha}\xspace}
\newcommand{\fekb}{\ensuremath{\mathrm{Fe}~\mathrm{K}\beta}\xspace}

\newcommand{\ergcms}{\ensuremath{\text{erg\,cm}^{-2}\,\text{s}^{-1}}\xspace}
\newcommand{\ergs}{\ensuremath{\text{erg}\,\text{s}^{-1}}\xspace}

\graphicspath{{./imgs/}}

\begin{document}

\title{\nustar observation of \gro at low mass accretion rate}

\author{Ole~K\"onig\inst{1} \and Felix~F\"urst\inst{2} \and
  Peter~Kretschmar\inst{2} \and Ralf~Ballhausen\inst{1} \and
  Ekaterina~Sokolova-Lapa\inst{1,3} \and Thomas~Dauser\inst{1} \and
  Celia~S\'anchez-Fern\'andez\inst{2} \and Paul~B.~Hemphill\inst{4}
  \and Michael~T.~Wolff\inst{5} \and Katja~Pottschmidt\inst{6,7} \and
  J\"orn~Wilms\inst{1}} \offprints{O.~K\"onig,
  \email{ole.koenig@fau.de}}

\institute{Dr. Karl-Remeis-Sternwarte and ECAP, Sternwartstr. 7, 96049
  Bamberg, Germany
  \and European Space Astronomy Centre (ESAC), Camino Bajo del
  Castillo, s/n., Urb. Villafranca del Castillo, 28692 Villanueva de
  la Ca{\~{n}}ada, Madrid, Spain
  \and Sternberg Astronomical Institute, M. V. Lomonosov Moscow State
  University, Universitetskij pr., 13, Moscow 119992, Russia
  \and MIT Kavli Institute for Astrophysics and Space Research, 77
  Massachusetts Ave, Cambridge, MA 02139, USA
  \and Space Science Division, Naval Research Laboratory, Washington DC 20375, USA
  \and CRESST, Department of Physics, and Center for Space Science and
  Technology, University of Maryland Baltimore County, 1000 Hilltop
  Circle,  Baltimore, MD 21250, USA
  \and NASA Goddard Space Flight Center, Code 661, 8800 Greenbelt Road, Greenbelt, MD 20771, USA
}

\abstract {Neutron stars in low-mass X-ray binaries (LMXBs) are
  important systems to study the physics of accretion onto compact
  objects. The system \gro is particularly interesting, as it usually
  shows clear pulsations as well as X-ray bursts. Additionally, there
  are claims for a magnetic field of $5\times 10^{11}$\,G through the
  detection of a Cyclotron Resonant Scattering Feature (CRSF).} {We
  present the spectral analysis of \gro using $\sim$29\,ks of \nustar
  data taken in 2017 February at a low luminosity of $3.2\times 10^{36}$\,\ergs (3--50\,keV). Our
  goal is to study the variability of the source spectrum with pulse
  phase and to search for the claimed CRSF.} {The continuum spectrum
  is modeled with an absorbed power-law with exponential cut-off, and
  an additional iron line component. We find no obvious indications
  for a CRSF and therefore perform a detailed cyclotron line search
  using statistical methods. We perform this search on pulse
  phase-averaged as well as phase-resolved spectra.} {\gro was
  observed in a low luminosity state. The previously detected Type~II
  X-ray bursts are absent. Clear pulsations at a period of
  2.141124(9)\,Hz are detected. The pulse profile shows an indication
  of a secondary peak, which was not seen at higher flux. The
  $4\,\sigma$ upper limit for the strength of a CRSF in the 3--20\,keV
  band is 0.07\,\kev, lower than the strength of the line found at
  higher luminosity.} {The detection of pulsations shows that the
  source did not enter the ``propeller'' regime, even though the
  source flux of $4.15\times 10^{-10}$\,\ergcms was almost one order of magnitude below the
  threshold for the propeller regime claimed in previous studies on
  this source. The transition into the propeller regime in \gro must
  therefore be below a luminosity of $3.2\times 10^{36}$\,\ergs (3--50\,keV), which implies a
  surface magnetic field $\lesssim 2.9\times 10^{11}$\,G and mass accretion rate
  $\lesssim 1.7\times 10^{16}$\,g\,s$^{-1}$. A change of the CRSF depth as function of
  luminosity is not unexpected and has been observed in other sources.
  This result possibly implies a change in emission geometry as
  function of mass accretion rate to reduce the depth of the line
  below our detection limit.}


\date{Received 29 July 2020 / Accepted 14 September 2020}

\maketitle

\section{Introduction}
\label{sec:intro}
Accretion-powered X-ray pulsars are binary systems consisting of a
neutron star and an optical companion. They are often classified by
the mass and spectral type of the donor star. High-mass X-ray binaries
\citep[HMXBs, e.g.,][and references
therein]{Chaty11a,paul:2011,reig:2011} have donor stars of O/B type.
The accretion is typically wind-fed or, in the case of Be/X-ray
binaries, from the Be star's decretion disk. The neutron star exhibits
a magnetic field strength of ${\sim}10^{12}$--${\sim}10^{13}$\,G).
Low-mass X-ray binaries \citep[LMXBs, e.g.,][]{Bhattacharyya10a}, on
the other hand, consist of a late-type donor star and accreting
neutron star, typically with lower magnetic field strength
\mbox{$\mathcal{O}$($10^8$--$10^9$\,G)}. These systems are believed to
be much older and have an accretion disk usually fed by Roche lobe
overflow.

Some pulsars exhibit spectral features which allow a direct estimate
of the neutron star's magnetic field. When ionized matter approaches
the Alfv\'en radius \citep{Alfven68}, it couples to the $B$-field
lines of the neutron star. The plasma is then funneled to the poles,
where it gets decelerated and forms so-called accretion columns.
Moreover, in the presence of a magnetic field the electrons' motion is
quantized perpendicular to the field onto discrete energy states, the
Landau levels \citep{Landau65,1981PhRvD..23..328L}. Inverse Compton
scattering -- which is mainly responsible for the hard X-ray radiation
-- becomes a resonant process in this case. The transition of the
electrons between different Landau levels gives rise to absorption
line-like features in the spectrum. These are referred to as cyclotron
resonant scattering features (CRSFs). The CRSF energy allows to infer
the $B$-field strength at the line forming region via the
``12-$B$-12'' rule \citep{CanutoVentura77},
\begin{equation}
  B\sim (1+z)(E_\mathrm{c}/11.6\,\kev)\times 10^{12}\,\mathrm{G}
\end{equation}
where $z$ is the surface gravitational red shift. CRSFs are usually
detected in HMXBs at 10--90\,\kev due to their intrinsic high
$B$-field strength. LMXBs typically do not exhibit these lines
although there are some famous exceptions \citep[e.g.,
  Her~X-1][]{1978ApJ...219L.105T}. A comprehensive review of CRSF
sources and their observation has been recently presented by
\citet{2019A&A...622A..61S}, while \citet{Schwarm17b,Schwarm17a}
summarize their theoretical modeling.

\gro is a transient LMXB discovered on 1995 December 2 with the
\textit{Burst And Transient Source Experiment} (BATSE) on-board the
\textit{Compton Gamma Ray Observatory}
\citep{Fishman95,kouveliotou96a}. It is associated with a position
near the Galactic center, at a distance of 7.5--8.5\,kpc
\citep{Augusteijn97A,Nishiuchi99A}, although smaller values have been
reported as well \citep{Sanna17a}. The companion star is of type G4
\textsc{iii} \citep{Gosling07A,2014ATel.5999....1M}, and has an
inferred mass of $<0.3\,\msun$ with inclination
$i>15^\circ$. \citet{Sanna17a} give the most recent orbital ephemeris,
with the orbital period $P_\mathrm{orb}=11.8358(5)$\,days, the
projected semi-major axis $a_x\sin(i)=2.639(1)$\,light-sec, and the
eccentricity $\epsilon<6\times 10^{-3}$. The source has a magnetic
field that is strong enough that X-ray pulsations at 2.14\,Hz are
observed \citep{1996Natur.381..291F}, even though Type~II X-ray bursts
are also seen \citep{Lewin76a}. \gro has therefore been dubbed the
``Bursting Pulsar'' \citep{Strickman96a}. Pure thermonuclear Type~I
X-ray bursts have not been detected, which is consistent with the
picture of a high magnetic field \citep{Bildsten97a,Court18a}.
\citet{Cui97} reports that the pulsations cease at a flux limit of
$2.3\times 10^{-9}\,\ergcms$ and interprets this threshold as a
``centrifugal barrier'' where the magnetosphere halts the accretion
flow. This is also known as the ``propeller'' effect. By equaling the
co-rotation radius and the radius of the magnetosphere
\citep[e.g.,][]{Fuerst17a}, one can estimate the surface magnetic field
from the luminosity threshold
\begin{equation}
  \mathrm{L}_\mathrm{prop}\simeq \frac{G M \dot{M}}{R}\simeq 7.3\times
  10^{37}~k^{7/2}~B_{12}^2~P^{-7/3}~M^{-2/3}_{1.4}~R^{5}_6~\ergs ~,
  \label{eq:bestimate}
\end{equation}
where $k=0.5$ in the case of disk accretion \citep{Ghosh78}, $B$ is
the magnetic field in units of $10^{12}$\,G, $P$ is the rotational
period of the neutron star, $M_{1.4}$ is its mass in units of
1.4\,\msun, and $R_6$ is its radius in units of $10^6$\,cm.

About one year after the 1995 outburst during which \gro was
discovered, it underwent a similar outburst in 1996 December
\citep{woods99a,doroshenko15a}, followed by 18\,years of quiescence.
In early 2014, \gro went into outburst again, which triggered \nustar
and \chandra \citep{younes15a} as well as \xmm and \integral
\citep{dai15a} observations. With $1.9\times10^{38}\,\ergs$ the peak
luminosity of the 2014 outburst was extremely high for a typical X-ray
binary, although still similar to the two earlier outbursts.

\begin{figure}
  \resizebox{\hsize}{!}{\includegraphics{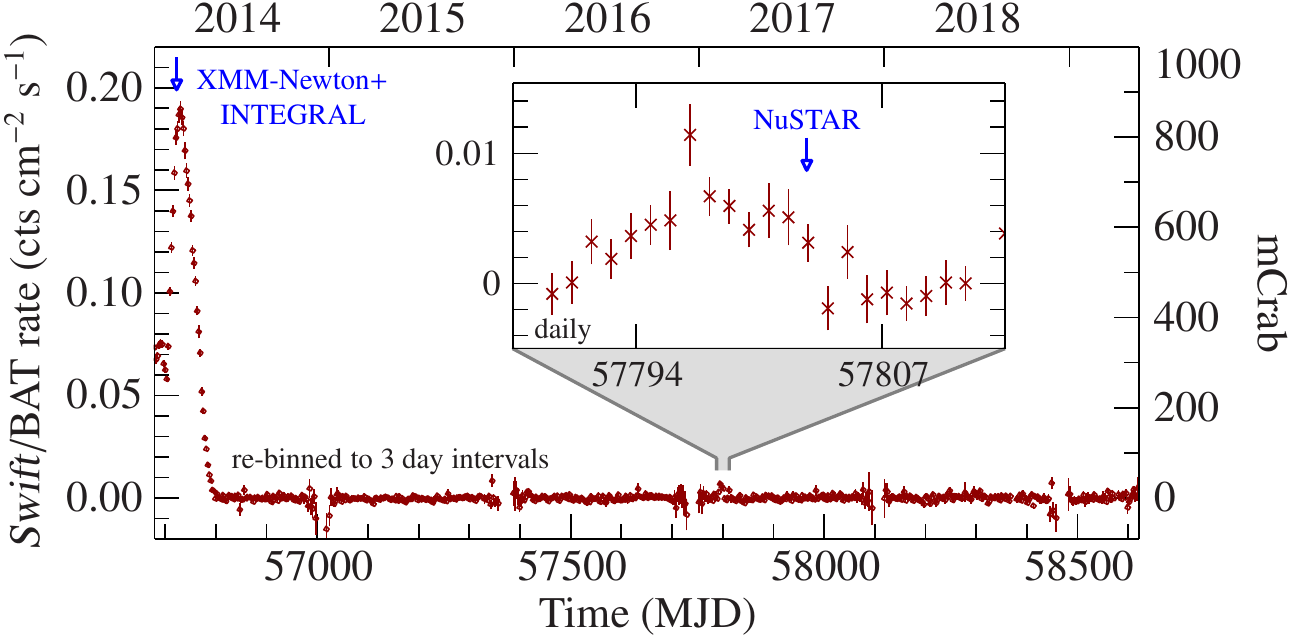}}
  \caption{\swift/BAT lightcurve of \gro. The outburst in 2014 had a
    peak luminosity of $2.1\times10^{38}\,\ergs$, slightly above the
    Eddington limit \citep{dai15a}. The outburst in 2017
    is at much lower luminosity $3.2\times 10^{36}$\,\ergs (3--50\,keV). The gaps in the
    \swift/BAT lightcurve are due to visibility constraints.}
  \label{fig:lc}
\end{figure}

This paper focuses on the fourth activity period of \gro, in 2017
February. The \swift/BAT \citep{Krimm13a} monitoring lightcurve is
shown in Fig.~\ref{fig:lc}, the overall luminosity of the source was
significantly less than that seen in previous outbursts. The pointed
\nustar observations during this outburst were performed at a flux of
$4.15\times 10^{-10}$\,\ergcms, which gives the first opportunity to study \gro in a regime of
low mass accretion rate.

Using \xmm and \integral data collected during the 2014 outburst,
\citet{dai15a} reported a fundamental CRSF at $4.68\pm0.05$\,\kev,
with an indication of second and third harmonics at $10.4\pm0.1$\,\kev
and $15.8^{+1.3}_{-0.7}$\,\kev. 
Shortly afterwards, \citet{doroshenko15a} claimed evidence for a CRSF
at $\sim4.5\,\kev$ in archival \sax data taken during the 1997
outburst. These claims make \gro one of the few LMXBs where a CRSF has
been reported, and one of the very few sources with a reported CRSF
energy below 10\,keV. Other pulsars with low CRSFs are
\object{4U\,1822$-$371} with a claimed energy of 0.7\,keV
\citep{2015A&A...577A..63I} and \object{Swift\,J0051.8$-$7320} at
5\,\kev \citep{2018MNRAS.480L.136M}.

The detection of the CRSF in \gro, however, is debated.
\citet{younes15a} did not find a significant CRSF in their data, which
were taken only three days earlier than the \citet{dai15a} detection.
The polar magnetic field deduced from the CRSF energy, $5\times
10^{11}$\,G, is higher than the ones derived with different methods
as, e.g., by \citet[2--6$\times 10^{10}\,\mathrm{G}$ from accretion
  disk reflection modeling]{degenaar14a}, \citet[$9\times
  10^{10}\,\mathrm{G}$ from a spin-up estimate]{younes15a}, and
\citet[$2.4\times 10^{11}\,\mathrm{G}$ from the propeller effect flux
  threshold]{Cui97}.

In this paper we discuss our analysis of \nustar observations of \gro
taken during its most recent outburst in 2017. In Sect.~\ref{sec:data}
we discuss the data extraction and calibration. We show that
pulsations are clearly present in the data and present the results of
the phase-averaged and phase-resolved spectroscopy in
Sect.~\ref{sec:phasavg} and~\ref{sec:phasres}, respectively, and
discuss the search for a CRSF. We discuss and summarize our results in
Sect.~\ref{sec:disc}.

\section{Data extraction \& calibration}
\label{sec:data}
The Nuclear Spectroscopic Telescope Array
\citep[\nustar;][]{Harrison13a} has an energy range from 3--79\,\kev
and temporal resolution of $2\,\mu\mathrm{s}$ which allows
phase-resolved spectroscopy of rapidly rotating neutron stars. It has
a moderate energy resolution of 400\,eV (FWHM) at 10\,\kev.

The \nustar data analyzed here have a net exposure of
28.8\,$\mathrm{ks}$ (FPM~A) and 28.9\,$\mathrm{ks}$ (FPM~B) starting
on 2017-02-18 14:34:35 UTC (MJD 57802.6073, ObsID: 80202027002),
during the decay of the outburst. After standard cleaning for Earth
occultation and the South Atlantic Anomaly according to the \nustar
data analysis software
guide\footnote{\url{https://heasarc.gsfc.nasa.gov/docs/nustar/analysis/nustar_swguide.pdf}},
we reduce the data with HEASOFT version~6.26 (corresponding to
NuSTARDAS 1.8.0), using \nustar CalDB version 20190513. We barycenter
the data and extract the source lightcurve and spectra from a circle
of 60\asec radius centered on the source position. We do an orbit
correction with the latest orbital parameters from \citet{Sanna17a},
although we note that the orbit is still poorly constrained and the
correction does not change the determined spin period significantly.
For the background extraction we define three circular regions of
120\,\asec radius for FPM~B and two circles of the same size for FPM~A
(due to stray light contamination). We then average the counts per
FPM~A/B and scale them to the source area to increase the background
statistics. The event files used to extract the phase-resolved spectra
were filtered on the source region with XSELECT version 2.4g.

All further analysis was performed with the \textit{Interactive
  Spectral Interpretation System} \citep[ISIS version
1.6.2-43]{2002hrxs.confE..17H}. Unless stated otherwise, all error
bars are at the 90\% level single parameter confidence level
($\Delta\chi^2=2.71$). We restrict the data to the 3--78\,\kev range
(PI channel 35--1210). We use the following binning scheme for the
phase-averaged and phase-resolved spectra to account for \nustar's
energy dependent energy resolution and oversample it by roughly a
factor of 3: In energy range 3--10\,\kev, we group a minimum number of
2 channels per bin, in range 10--15\,\kev:~3 channels,
15--20\,\kev:~5, 20--35\,\kev:~8, 35--45\,\kev:~16, 45--55\,\kev:~18,
55--65\,\kev:~48, 65--76\,\kev:~72, and in the range $>$76\,\kev, we
bin to 48 channels per bin, while also ensuring a minimum
signal-to-noise ratio (\snr) of 5.

\section{Spectral and Timing Analysis of GRO\,J1744$-$28}
\subsection{Phase-averaged spectrum}
\label{sec:phasavg}

\begin{figure*}
  \resizebox{\hsize}{!}{\includegraphics{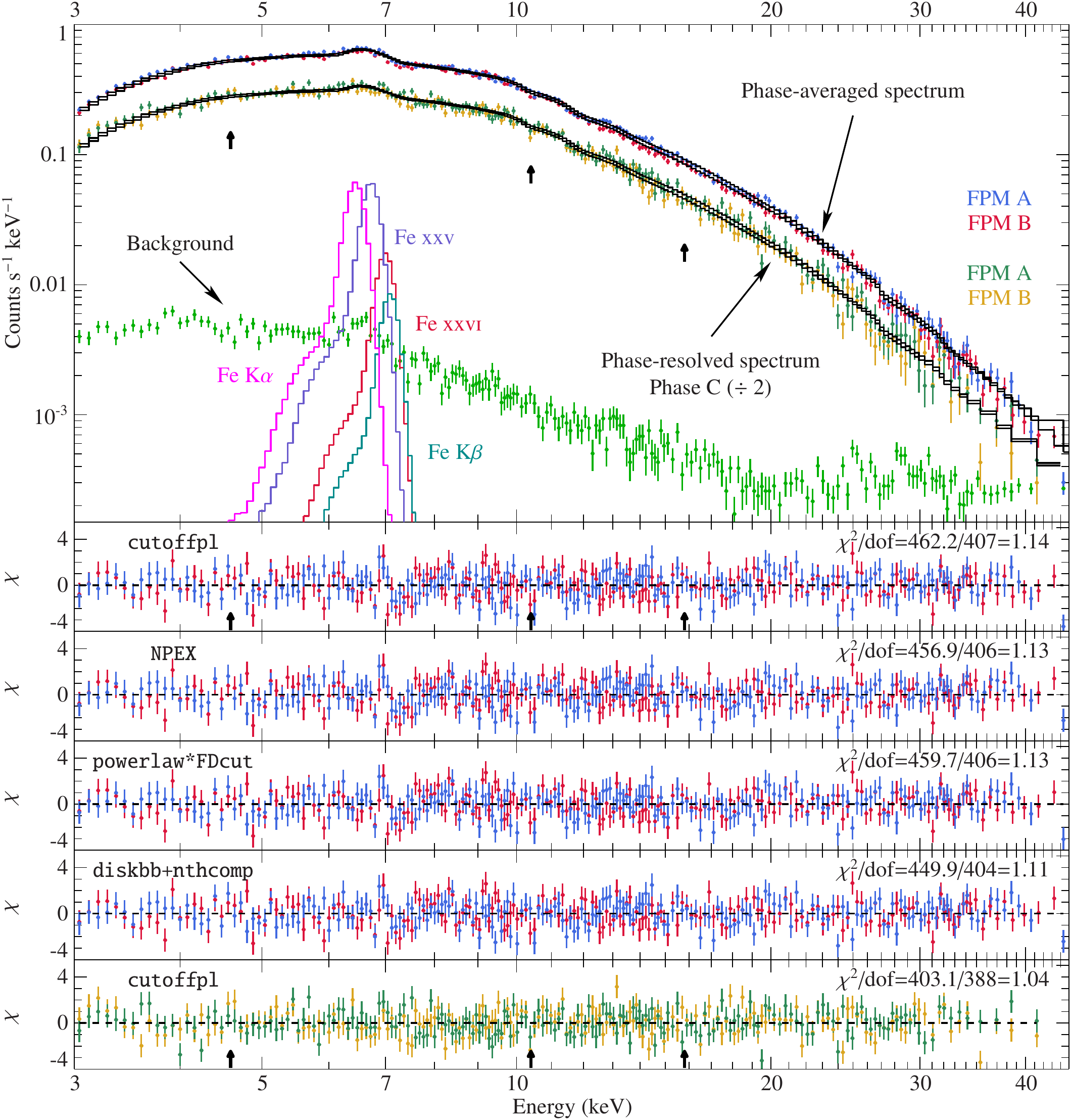}}
  \caption{Phase-averaged and one phase-resolved (phase C as
    defined in Fig.~\ref{fig:HR_PP}, displaced by factor of 2 for
    visualization) spectrum.
    Black histogram gives the best fitting model: an absorbed cut-off
    power-law with iron component at $\sim$6.5\,\kev.
    Green points show background of FPM~A.
    Arrows indicate the location of the reported CRSFs
    \citep{doroshenko15a,dai15a}.
    The iron line asymmetry is due to the convolution with the
    detector response and the logarithmic scale.
    All tested continuum models describe the data comparably well.
  }
  \label{fig:spectrum}
\end{figure*}

In order to allow us to compare the continuum shape with earlier
analyses, we use phenomenological continuum models rather than more
physically motivated models such as those by
\citet{beckerwolff07}\footnote{The \citet{beckerwolff07} model is also
  not applicable because the luminosity is too low to assume the
  presence of a radiation-dominated radiative shock.} or
\citet{Farinelli16a}. As discussed, e.g., by
\citet{2013A&A...551A...6M}, phenomenological spectral models
typically used to describe the continua of accreting neutron stars are
the exponentially cut-off power-law (\texttt{cutoffpl}), the power-law
with Fermi-Dirac cut-off \citep[\fdcut]{1986LNP...255..198T}, a
negative-positive cut-off power-law
\citep[\npex]{1995PhDT.......215M}, and a model consisting of a
black-body disk \citep[\texttt{diskbb}]{1984PASJ...36..741M} and
thermally comptonized continuum
\citep[\texttt{nthcomp}]{1996MNRAS.283..193Z,1999MNRAS.309..561Z}. The
residuals of the \cutoffpl, \fdcut, \npex and \texttt{diskbb+nthcomp}
models are shown in Fig.~\ref{fig:spectrum} and the best fit
parameters are given in Table~\ref{tab:fitparams}. The \npex and
\fdcut residuals look very similar, because they are driven to
parameters which effectively mimic the \cutoffpl solution. All tested
continuum models describe the data similarly well. Due to its
simplicity and in order to allow comparison with previous work
\citep[e.g.,][]{younes15a}, we use the \cutoffpl model for all
subsequent analysis. Photoelectric absorption in the interstellar
medium is accounted for with the \texttt{tbnew} model (\texttt{TBabs
  in \textsl{XSPEC}}) with cross sections and abundances according to
\citet{1996ApJ...465..487V} and \citet{2000ApJ...542..914W},
respectively. The iron fluorescence line complex can formally be
described by a slightly broadened ($\sigma=0.23^{+0.05}_{-0.04}$\,keV)
Gaussian component at $6.59\pm0.04$\,keV. This is most likely a blend
of different ionization states that cannot be resolved with \nustar.
Strongest fluorescence lines are often produced by neutral (6.4\,keV),
He- (6.7\,keV), and H-like iron (7.0\,keV), and the structure seen in
the data is also consistent with a set of narrow K$\alpha$ lines from
these ions, as well as neutral K$\beta$ (7.1\,keV) with a
K$\beta$/K$\alpha$ flux ratio of 13\%
\citep{2003A&A...410..359P}. With fixed energies and widths, this
approach is also statistically valid and has the same degrees of
freedom as using one broad emission feature but shows less
interference with the continuum modeling because all line energies and
widths are fixed and broadening is only due to the detector
response. Using both approaches, slight residuals still remain at the
iron K edge. These residuals are due to a combination of a gain-shift in
\nustar energy calibration and the fact that the \texttt{tbnew} model
only includes neutral iron.

The full model used for the X-ray continuum in our spectral fits with
ISIS is therefore
\begin{equation}
N_\mathrm{ph}(E) = \mathtt{tbnew}*\mathtt{const}*(\mathtt{cutoffpl}+\mbox{Fe-complex}) .
\end{equation}
where the iron complex is modeled with Gaussian emission lines and the
constant accounts for potential flux calibration uncertainties between
FPM~A and FPM~B. This model gives a good description of the overall
continuum shape ($\chi^2/\mathrm{dof}=462.2/407=1.14$). The fit
statistics of the best-fit \fdcut, \npex, and \texttt{diskbb+nthcomp}
models are shown in Table~\ref{tab:fitparams}. The observed
3--50\,\kev flux of $4.15\times 10^{-10}$\,\ergcms translates to a luminosity of $3.2\times 10^{36}$\,\ergs (3--50\,keV),
assuming spherical emission and a distance of 8\,kpc. This is roughly
two orders of magnitude lower than in the 1997 and 2014 outbursts, but
three orders of magnitude brighter than the quiescent detections
discussed by \citet{Daigne02a} and \citet{WijnandsWang02a} who see
a softer spectrum.

\begin{table*}
  \caption{Fit parameters of phase-averaged spectrum for various models}
  \label{tab:fitparams}
    \begin{tabular}{lllll}
      \hline\hline
      Parameter & \texttt{cutoffpl} & \npex & \texttt{pl}$\times$\fdcut & \texttt{diskbb+nthcomp} \\
      \hline
      $N_\mathrm{H} \:(10^{22}\,\text{cm}^{-2})$ & $4.9\pm0.6$ & $4.5\pm0.6$ & $5.9\pm0.6$ & $10\pm4$ \\
      $\Gamma$ & $0.53\pm0.05$ & $0.25\pm0.04; -2\tablefootmark{a}$ & $0.83^{+0.08}_{-0.04}$ & $1.81^{+0.06}_{-0.05}$ \\
      $E_{\text{fold}}\: (\kev)$ & $9.03^{+0.24}_{-0.23}$ & $5.56^{+0.24}_{-0.20}$ & $9.24\pm0.21$ & \\
      $E_{\text{cut}}\: (\kev)$ & & & $\le3.19$ & \\
      Flux ($\times 10^{-10}$\,\ergcms, 3--50\,\kev) & 4.153$\pm$0.027
      & 4.139$\pm$0.026 & 4.144$\pm$0.026 & $4.140^{+0.028}_{-0.027}$\\
      \hline
      $kT_\mathrm{e}\: (\kev)$ & & & & $5.99^{+0.25}_{-0.22}$ \\
      $kT_\mathrm{BB}\: (\kev)$ & & & & $1.41^{+0.20}_{-0.12}$ \\
      $kT_{\text{disk}}\: (\kev)$ & & & & $0.71^{+0.29}_{-0.12}$ \\
      $\text{Norm}_{\texttt{diskbb}}$ & & & & $30^{+120}_{-40}$ \\
      \hline
      $E_{\feka}\:(\kev)$\tablefootmark{ac} & 6.404 & 6.404 & 6.404 & 6.404 \\
      Flux\:($10^{-4}$\,ph s$^{-1}$cm$^{-2}$) & $1.23\pm0.26$ & $1.20\pm0.26$ & $1.25\pm0.26$ & $1.12^{+0.30}_{-0.31}$ \\
      $E_{\fekb}\:(\kev)$\tablefootmark{ac} & 7.058 & 7.058 & 7.058 & 7.058 \\
      Flux\:($10^{-5}$\,ph s$^{-1}$cm$^{-2}$)\tablefootmark{b} & $1.6$ & $1.5$ & $1.6$ & $1.4$ \\
      $E_{\ion{Fe}{xxv}}\:(\kev)$\tablefootmark{ac} & 6.7 & 6.7 & 6.7 & 6.7 \\
      Flux\:($10^{-4}$\,ph s$^{-1}$cm$^{-2}$) & $1.20\pm0.28$ & $1.20\pm0.28$ & $1.19\pm0.28$ & $1.11^{+0.29}_{-0.30}$ \\
      $E_{\ion{Fe}{xxvi}}\:(\kev)$\tablefootmark{ac} & 6.98 & 6.98 & 6.98 & 6.98 \\
      Flux\:($10^{-5}$\,ph s$^{-1}$cm$^{-2}$) & $3.3\pm2.3$ & $3.2\pm2.3$ & $3.8\pm2.3$ & $2.8^{+2.5}_{-2.6}$ \\
      \hline
      $E_\text{gabs}\:(\kev)$ & $7^{+13}_{-4}$ & & & \\
      $\sigma_\text{gabs}\:(\kev)$\tablefootmark{a} & 1.0 & & & \\
      Strength (\kev) & $\le0.07$ & & & \\
      \hline
      Normalization & $0.0141\pm0.0010$ & \begin{tabular}{l}$0.0120^{+0.0010}_{-0.0009}$ \\ $\left(1.21^{+0.27}_{-0.28}\right)\times10^{-3}$\end{tabular} & $0.0364^{+0.0025}_{-0.0037}$ & $\left(3.3^{+0.8}_{-0.9}\right)\times10^{-3}$ \\
      Constant & $0.965\pm0.007$ & $0.965\pm0.007$ & $0.965\pm0.007$ & $0.965\pm0.007$ \\
      $\chi^2$\,(dof) & 462.2/407 & 456.9/406 & 459.7/406 & 449.9/404 \\
      $\chi^2_{\text{red.}}$ & 1.14 & 1.13 & 1.13 & 1.11 \\
      \hline
    \end{tabular}%
  \tablefoot{
    The underlying function for all fit models is
    $\mathtt{tbnew}*\mathtt{const}*(\mathtt{\bf continuum~model}+\mbox{Fe-complex})$ where the
    continuum model is specified in the column header.
    The upper limit on a cyclotron line is discussed in
    Sect.~\ref{sec:phasavg}.
    Uncertainties are at the 90\% confidence level. We use
    \texttt{wilm} abundances and \texttt{vern} cross-sections.\\
    \tablefoottext{a}{Parameter frozen}
    \tablefoottext{b}{Tied to $0.13\cdot\text{Flux}_{\feka}$, see
      \citet{2003A&A...410..359P}}
    \tablefoottext{c}{Narrow line with frozen width
      $\sigma=10^{-6}\,\kev$}
  }
\end{table*}

\subsection{Phase-resolved spectra}
\label{sec:phasres}

\subsubsection{Pulse period}
As the neutron star rotates, the line of sight onto the accretion
column changes. Most X-ray pulsars therefore show spectral variations
as a function of phase \citep[see, e.g.,][]{2011A&A...532A..76F}. To
extract phase-resolved spectra we first identify the local pulse
period and define phase-bins according to the hardness ratio, as
described below. Using the epoch folding technique
\citep{1983ApJ...272..256L,1989MNRAS.241..153S}, we find a pulse
period of 0.4670444(20)\,s, corresponding to a rotational frequency of
2.141124(9)\,Hz. The uncertainty is conservatively estimated by
$\Delta P = P^2/(2T_\mathrm{elapse})$. The determined pulse period is
consistent with earlier measurements
\citep{doroshenko15a,younes15a,dai15a}, indicating that little spin-up
of the neutron star has happened after the 2014 outburst.

\subsubsection{Pulse profile and continuum parameters}
\begin{figure}
  \resizebox{\hsize}{!}{\includegraphics{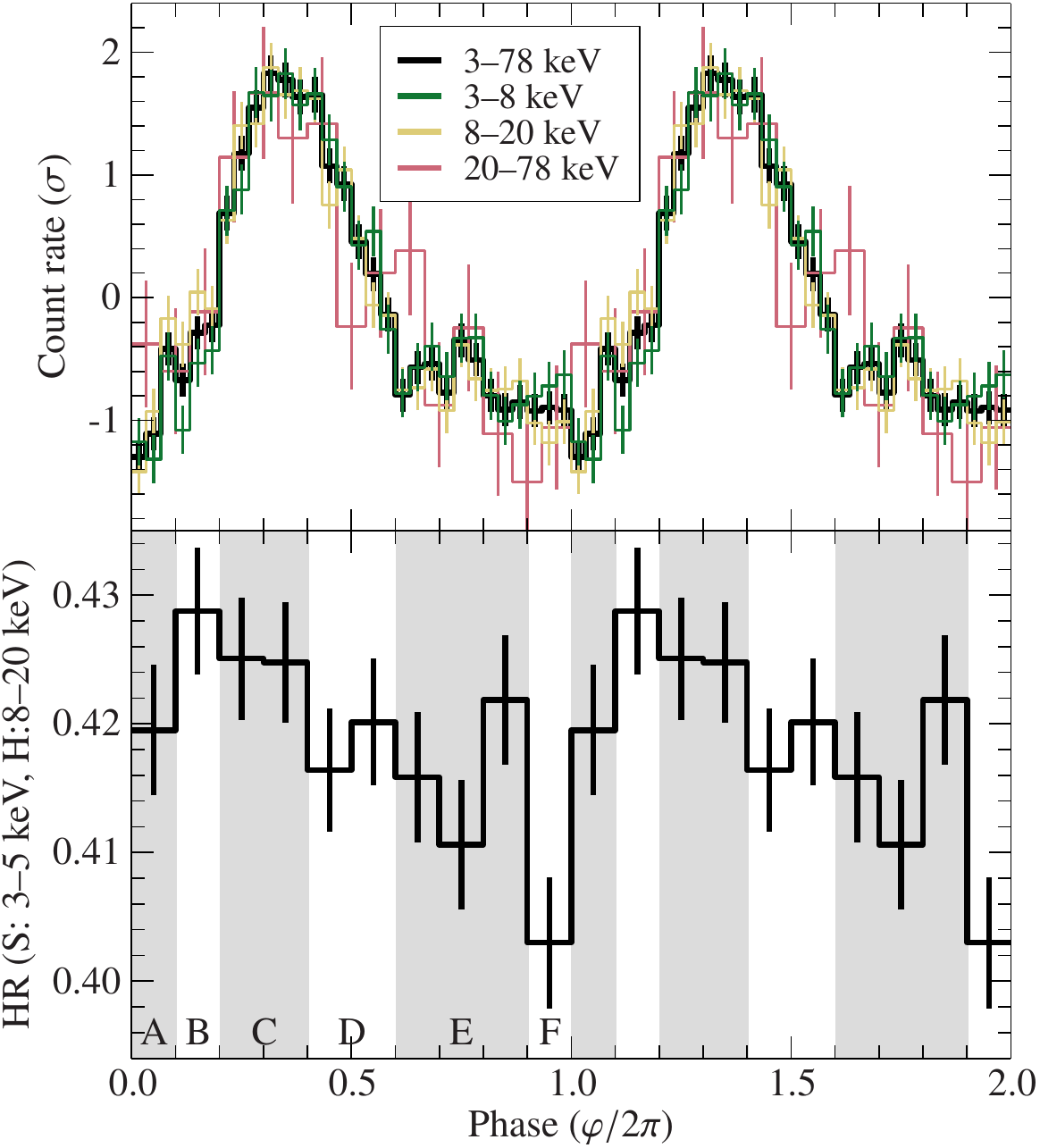}}
  \caption{\textit{Top:} Background-subtracted and GTI-corrected pulse
    profile of \gro for the 3--78\,keV band (black) and for three
    narrower energy bands (colored). The count rate is normalized by
    subtracting the mean and dividing by the standard deviation to
    emphasize potential changes in the shape of the profile. The
    20--78\,\kev band is binned more coarsely for visualization
    purposes. No significant energy dependence can be seen (see text).
    \textit{Bottom:} Variation of the hardness ratio with phase
    (combined data from FPM~A and B). The vertical bands show the
    phase ranges chosen for phase-resolved spectroscopy. }
  \label{fig:HR_PP}
\end{figure}

The top panel of Fig.~\ref{fig:HR_PP} shows the pulse profile obtained
by folding the energy-resolved lightcurves with the local pulse
period. The profile was cleaned for the Good Time Intervals of the
observation and background subtracted. It shows one prominent peak
spanning $\sim$1/3 of the rotation, and a decrease in flux to a
plateau at late phases where a secondary peak seems to be present
around phase 0.75. The colored histograms show the pulse profile in
different energy bands. There are only subtle changes in the pulse
spectral shape -- a KS-test \citep{Kolmogorov33,Smirnov39} yields no
significant energy dependence. The chances for belonging to the same
probability distribution is 96\% when comparing the 3--8\kev and
8--20\,\kev pulse profiles, 59\% for 3--8\kev vs.\ 20--78\,\kev, and
81\% for 8--20\,\kev vs.\ 20--78\,\kev.

\begin{figure}
  \resizebox{\hsize}{!}{\includegraphics{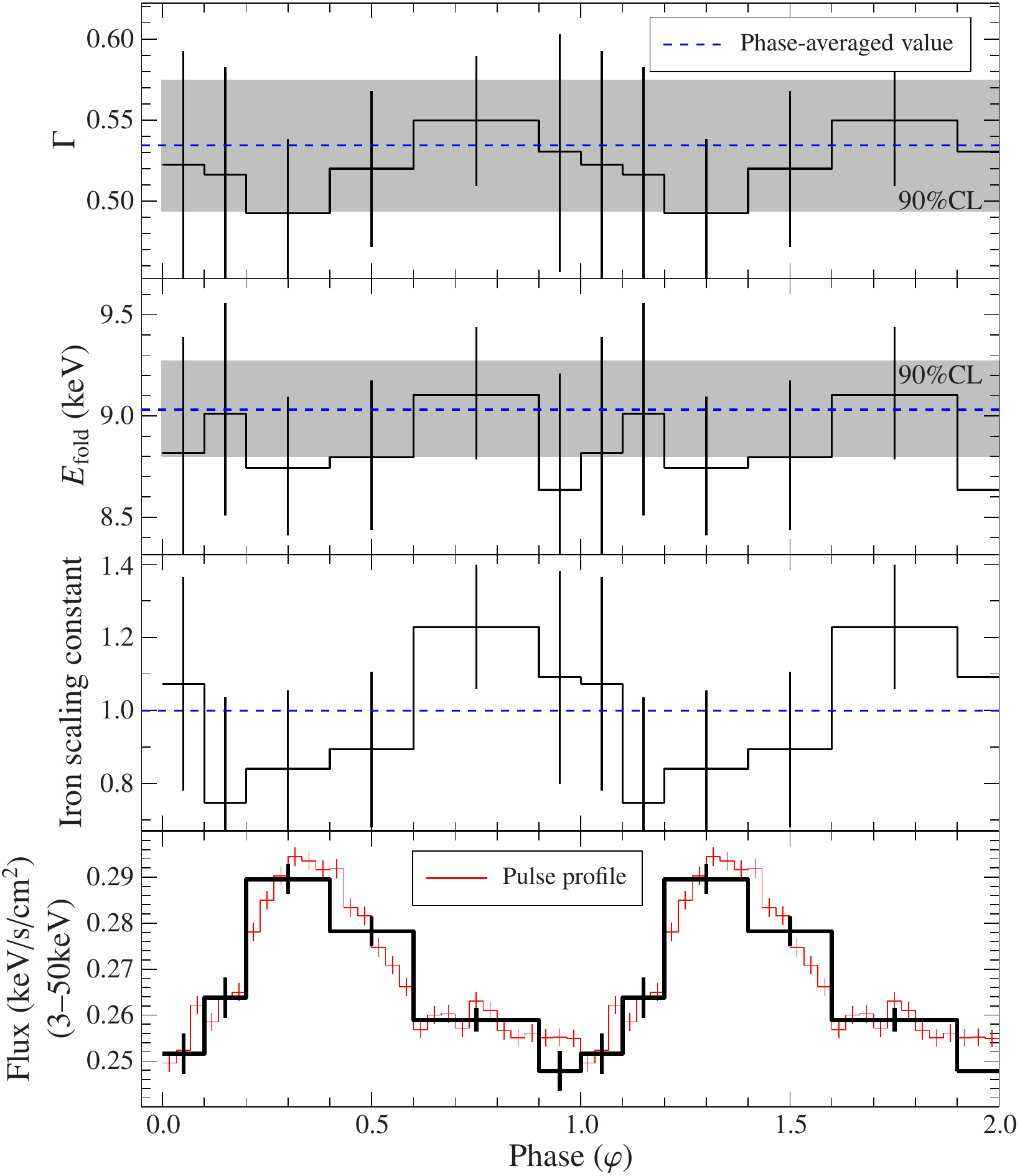}}
  \caption{Phase-resolved continuum parameters of the fit model. Blue
    dashed line display the phase-averaged values with 90\% confidence
    level (shaded). The FPM~A and FPM~B cross-calibration constant
    0.965 and the column density $6.35\times 10^{22}\,\text{cm}^{-2}$
    are fixed to the phase-averaged best fit values.  The ``iron
    scaling constant'' shows the variation of the relative strength of
    the iron line complex, whose parameters have been otherwise fixed
    to the phase-averaged values.  No significant changes can be
    identified in the continuum parameters.  In the lowest panel we
    plot the (re-scaled) pulse profile.  Errors are at the 90\%
    confidence level.}
  \label{fig:continuum}
\end{figure}

To study the spectral variability further, we calculate the hardness
ratios as \mbox{$(\mathrm{h}-\mathrm{s})/(\mathrm{h}+\mathrm{s})$},
where $h$ is the count rate in the hard band, and $s$ the count rate
in the soft band \citep{Lightman79}. To define bins for phase-resolved
analysis we take the hardness ratio between the 3--5\,keV (soft) and
8-20\,keV (hard) band into account, which provides a good compromise
between energy resolution and signal-to-noise ratio. To study the
slight variations present in the hardness, we define six phase bins
(A--F) of variable length, to cover periods of largely constant
hardness ratio. Since the different continuum parameters yielded
similar description of the phase-averaged continuum, we model the
phase-resolved data with the \cutoffpl model only. No large changes as
a function of phase are seen (Fig.~\ref{fig:continuum}), as expected
from the near constant hardness ratio.

\begin{figure}
  \resizebox{\hsize}{!}{\includegraphics{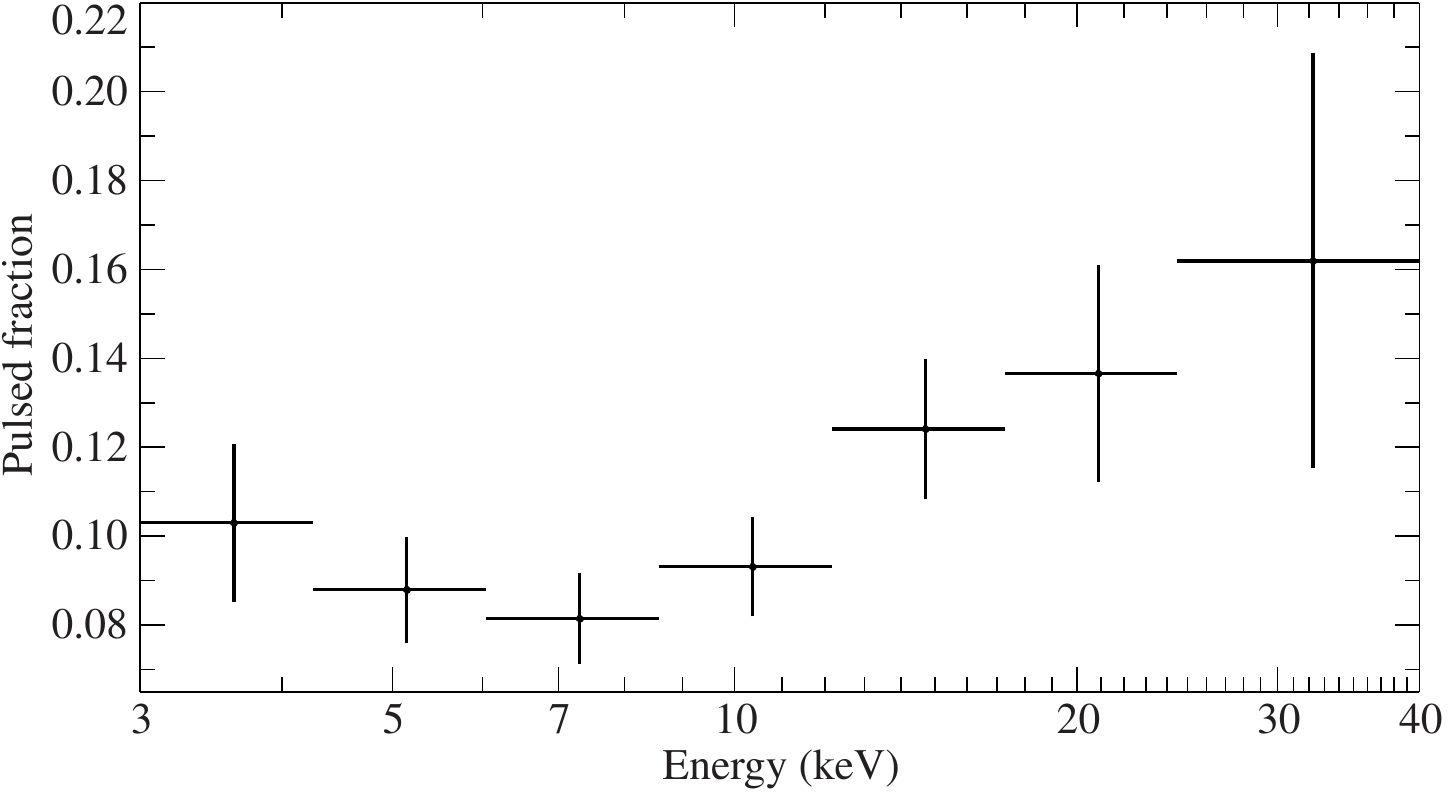}}
  \caption{Pulsed fraction
    $(\mathrm{CR}_\mathrm{max}-\mathrm{CR}_\mathrm{min})/(\mathrm{CR}_\mathrm{max}+\mathrm{CR}_\mathrm{min})$.
    The dip at 7\,\kev is most likely due to the iron fluorescence
    line. The increase at higher energies is consistent with previous
    findings.}
  \label{fig:pulsedfraction}
\end{figure}

Finally, we turn to the amplitude of the pulsation, as shown in
Fig.~\ref{fig:pulsedfraction}, as the energy dependent pulsed
fraction, defined here as
$(\mathrm{CR}_\mathrm{max}-\mathrm{CR}_\mathrm{min})/(\mathrm{CR}_\mathrm{max}+\mathrm{CR}_\mathrm{min})$,
where $\mathrm{CR}$ abbreviates the count rate in the energy band's
pulse profile. We emphasize that binning and the quality of the pulse
profiles can have an effect on the measurement of the pulsed fraction,
and that it is only a proxy for assessing the pulsations. The pulsed
fraction is 9.5$\pm$1.1\% at energies 3--6\,keV and 12.9$\pm$2.8\%
above 9\,keV, consistent with \citet{younes15a}. We see a tentative
dip at 7\,keV where the iron line is located. This is consistent with
the picture that this fluorescence line originates from outside of the
accretion column/hot-spot. The pulsed fraction over the full
3--78\,keV range is 8.2$\pm$0.6. We note that increases at high
energies have also been seen in other X-ray pulsars
\citep[e.g.,][]{Lutovinov09a}.

\subsection{Search for cyclotron resonant scattering feature}
\label{susec:crsf_search}

\subsubsection{Cyclotron line search in phase-averaged spectrum}
\label{sususec:crsf_phavg}

As discussed in Sect.~\ref{sec:intro}, \gro is among the CRSF
candidates with the lowest line energies proposed so far. A detailed
search in this \nustar observation at a luminosity around two orders of
magnitude lower than in previous outbursts is therefore of
particular interest. Since the residuals of our best-fit model without
a CRSF (Fig.~\ref{fig:spectrum}) do not show absorption line-like
residuals, the CRSF in the present observation must be weak or absent.
We therefore perform a systematic search for a line in order to at
least find limits for its parameters. The most common phenomenological
model is the multiplicative Gaussian absorption line (\gabs) of the form
$\exp[-\tau(E)]$ with the Gaussian-shaped optical depth
\begin{equation}
  \label{eq_tau}
  \tau(E)=\tau_0 \exp \left[-\frac{(E-E_c)^2}{2\sigma^2}\right]
\end{equation}
and $\tau_0$ the central optical depth. This component introduces the
parameter ``strength'' (in \kev) equaling $\tau_0\sqrt{2\pi}\sigma$
which is widely used in order to determine the significance of
cyclotron lines \citep[e.g.,][]{Pottschmidt05a,Lutovinov17a}. The
width of the line, $\sigma$, is mainly constrained by the electron
temperature, and the viewing angle
\citep[e.g.,][]{Heindl04a,Schwarm17a}. Based on \citet{Meszaros85a},
\citet{2019A&A...622A..61S} predict a width of $\sim$1\,keV at the
reported 5\,keV energy. Indeed, \citet{doroshenko15a} found a width of
$1.2\pm0.3$\,keV whereas \citet{dai15a} found a smaller width of
$0.68\pm0.08$\,keV. The weakness of any line in the spectrum will make
it impossible to constrain both, its energy and width. Motivated by
these earlier observations we therefore fix its width to 1\,keV and
perform a systematic search by stepping through the 3--20\,keV band in
1000 steps, while fitting for the line. This, however, does not give
improvement in $\chi^2$. The data yield an upper limit of 0.07\,\kev
(90\% CL) on the strength of a \gabs component. Attempts to model the
spectrum with a CRSF with varying width did not result in physical
values.

\subsubsection{Cyclotron line search in phase-resolved spectra}
\label{sususec:crsf_phres}

Although no significant cyclotron line is found in the phase-averaged
spectrum, it might still be possible that the line is present in
phase-resolved data. The reason is that CRSF produced in localized
regions in the accretion column might only be visible during certain
phases of the neutron star rotation.

\begin{figure}
  \resizebox{\hsize}{!}{\includegraphics{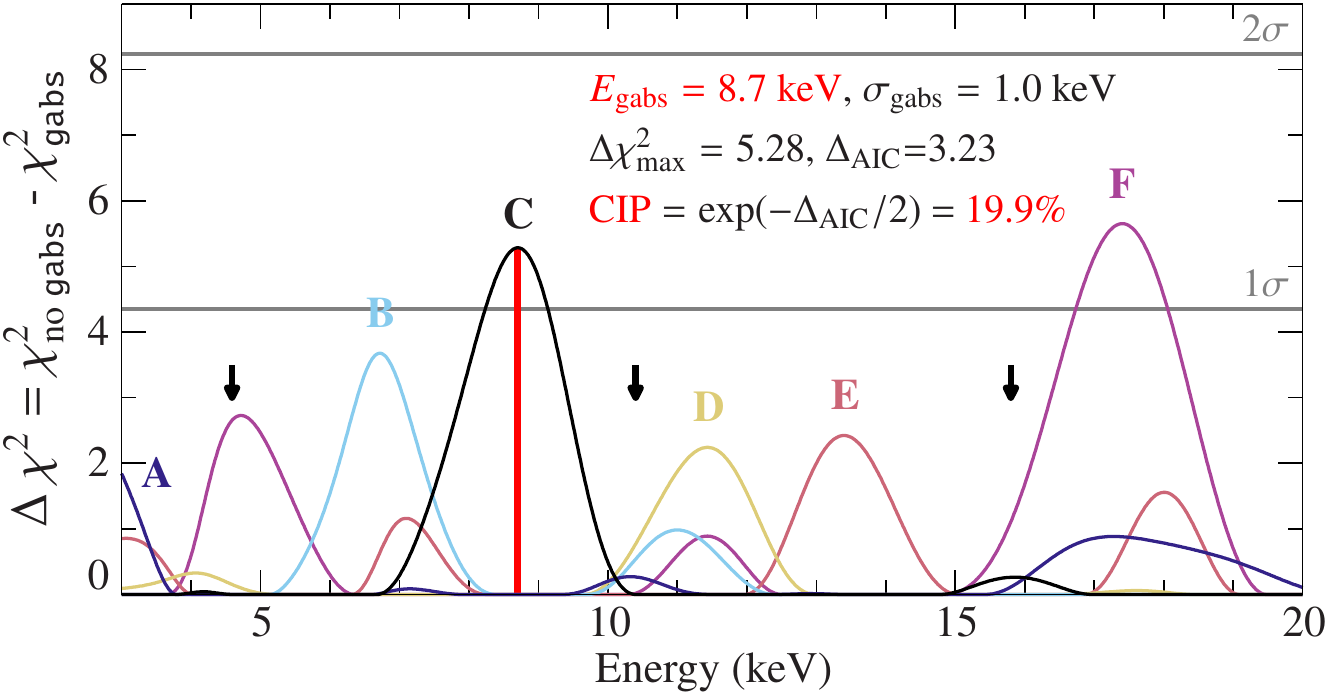}}
  \caption{$\Delta \chi^2$ for inclusion of a CRSF plotted as a
    function of energy for all phase-resolved spectra. No phase shows
    an absorption feature with $>2\sigma$ significance. Black arrows
    show the reported CRSF energies \citep{doroshenko15a,dai15a}.}
  \label{fig:AIC}
\end{figure}

To search for such a line, similar to our phase-averaged spectral
analysis, we include a Gaussian absorption component of fixed energy
and width and only fit the line strength. Based on the earlier data
summarized in Sect.~\ref{sec:intro}, we vary the centroid energy from
3\,keV to 20\,keV and determine the $\chi^2$-improvement at each
sampled energy. We again fix the line width to 1\,\kev \citep[see
  also][Fig.~12]{2019A&A...622A..61S}. We repeat this procedure for
all phase intervals and show the result in Fig.~\ref{fig:AIC}. Since
one cannot use simple likelihood ratio tests for the presence of a
line \citep{Protassov02a}, in order to see whether there are
significant deviations from the model without absorption line, we use
the \textit{Akaike Information Criterion}
\citep{1974ITAC...19..716A}. For small sample sizes, this is computed
by $\mathrm{AIC}=\chi+2k+(2k^{2}+2k)/(n-k-1)$ where $k$ is the number
of free parameters and $n$ the number of bins ($n-k$ is the number of
degrees of freedom). We cannot find any significant deviation in the
resulting \mbox{$\Delta\chi^2$-distribution} of Fig.~\ref{fig:AIC}: No
phase exhibits a significance larger than $2\sigma$. Specifically, the
largest $\Delta\chi^2$ in phase C yields a \emph{Chance
  Improvement Probability}, $\exp(-\Delta_{\mathrm{AIC}}/2)$, of 20\%.
It is therefore very likely that the slight increase of $\chi^2$ found
when including a CRSF is only due to statistical effects.

\subsubsection{Monte-Carlo simulations}
\label{sususec:MC}

\begin{figure}
  \resizebox{\hsize}{!}{\includegraphics{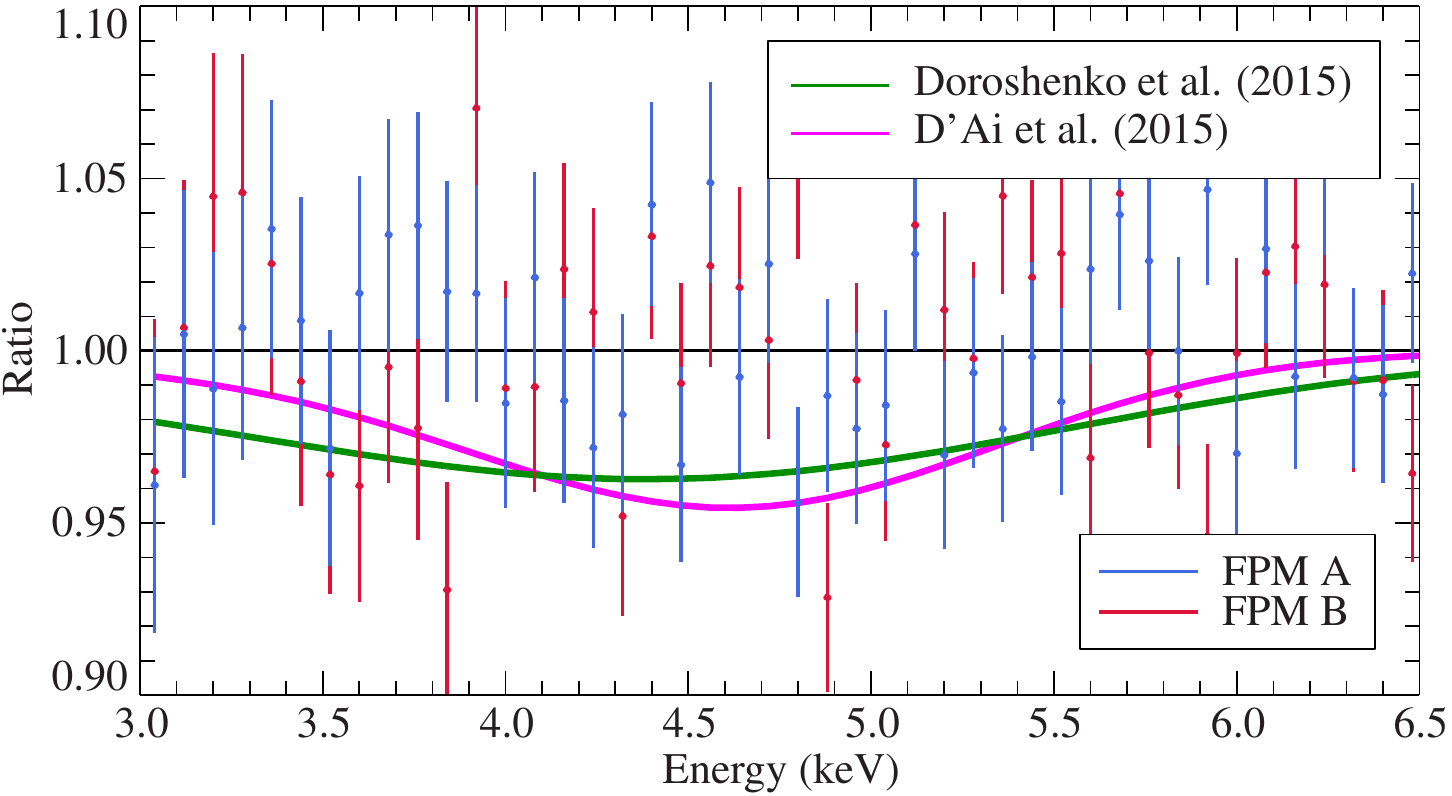}}
  \caption{Ratio of \nustar data and fit model without \gabs as
    function of energy in the band where a CRSF was observed in
    earlier data. The green and pink lines show the CRSF parameters
    claimed in earlier analyses. CRSFs with these strengths would have
    likely been seen in the present data data.}
  \label{fig:ratio}
\end{figure}

In Sect.~\ref{sususec:crsf_phavg}, we determined an upper limit of
0.07\,\kev on the strength of a CRSF. As the \gabs strength of
\citet{dai15a} and \citet{doroshenko15a} is above this value (with
0.087\,\kev and $\sim$0.12\,\kev, respectively), we can rule out a
line as strong as previously claimed at the 90\% level. We therefore
would likely have seen a trace (with $\sim$2--3$\sigma$) of the CRSF
if it was as strong as previously reported. This is illustrated in
Fig.~\ref{fig:ratio} where we plot the CRSF properties of previous
claims into the residuals of our \nustar data.

In this section we discuss how strong the cyclotron line would have to
be in order to be significantly detected in our \nustar data. We
simulate 20\,000 fake spectra based on the exposure and best fit model
of the phase-averaged spectrum (without \gabs). The data points are
drawn from a Poisson-distribution with mean at the model value. The
\snr-ratio of the data directly affects the amount of absorption
features emerging due to statistical fluctuations. By analyzing the
number of spurious detections of lines we put a lower limit on the
cyclotron line strength and determine whether we would have been able
to see the CRSF as previously reported.

\begin{figure}
  \resizebox{\hsize}{!}{\includegraphics{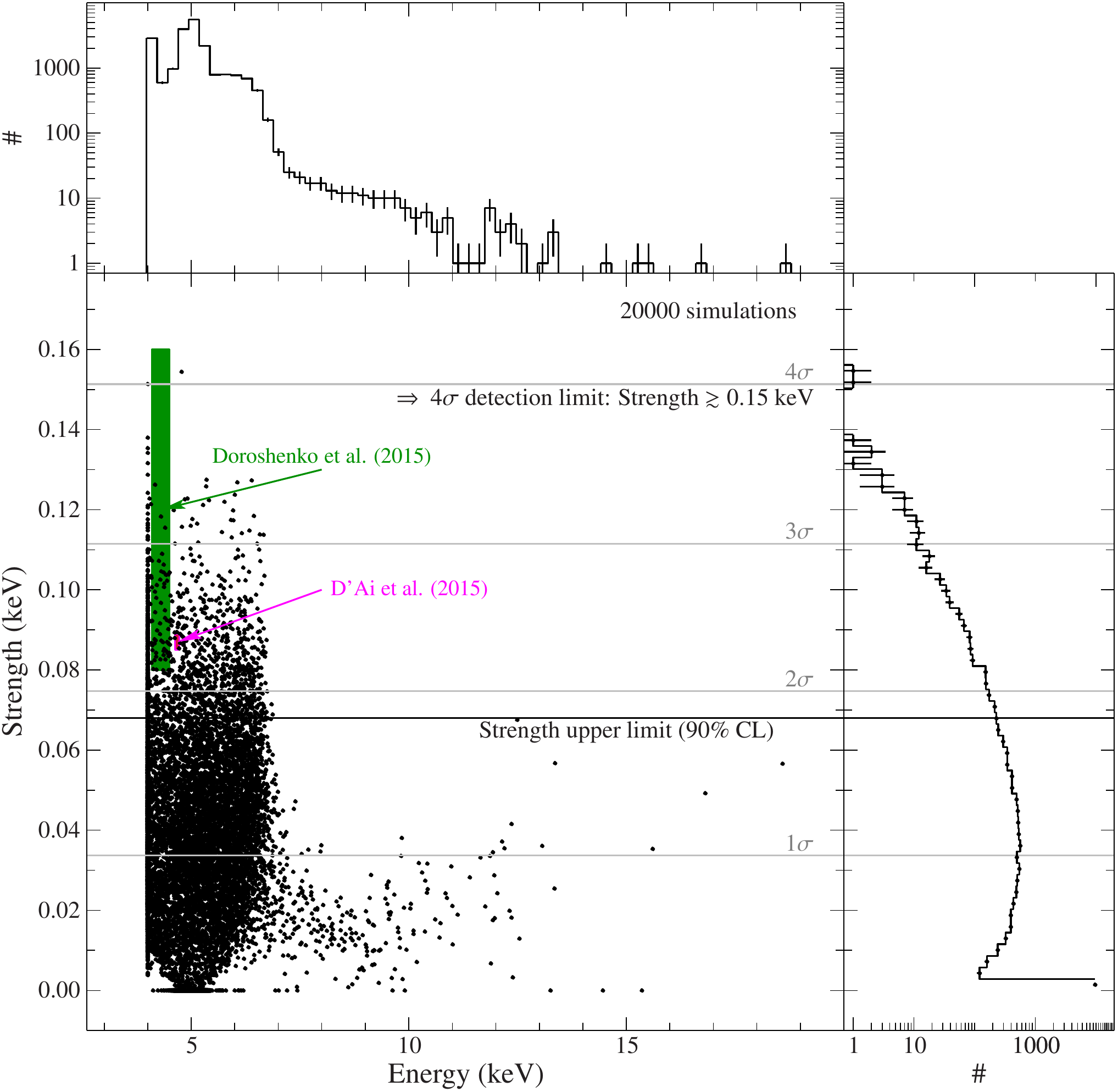}}
  \caption{Distribution of energy and strength of the best-fit CRSF
    found in a Monte Carlo Simulation of 20\,000 fake spectra that do
    not include a CRSF. Statistical fluctuations will lead to
    artificial absorption features, fitted with a \gabs model. We
    determine a $4\sigma$ detection limit of 0.15\,keV on the strength
    of a significant CRSF detection. 7\% of the fits are above the
    phase-averaged 90\% CL upper limit strength of 0.07\,\kev. }
  \label{fig:tau}
\end{figure}

We fit the simulated spectrum with the best phase-averaged fit model
plus an additional Gaussian absorption feature (\gabs) and extract its
strength and energy (Fig.~\ref{fig:tau}). We constrain the fitted
line energy of to be above 4\,\kev in order to avoid the line running
into \nustar's lower energy limit. The starting value of the line
energy is 5\,\kev.  Additionally, we freeze the width to 1\,keV as
before -- otherwise the width is almost always fitted to the lowest
possible value.
The Monte Carlo simulations of Fig.~\ref{fig:tau} show that many
spurious lines are above the values of previous reports. For a robust
detection claim, we would like to detect a line with a $4\sigma$
confidence level. In order to obtain how strong such a cyclotron line
must be in our \nustar data, we scan the ``\gabs strength''
distribution for the value where it exceeds the required false rate of
$6.3\times 10^{-5}$ ($4\sigma$). Thus, we put a $4\sigma$ detection
limit of 0.15\,keV on the strength of the cyclotron line.

\begin{figure}
  \resizebox{\hsize}{!}{\includegraphics{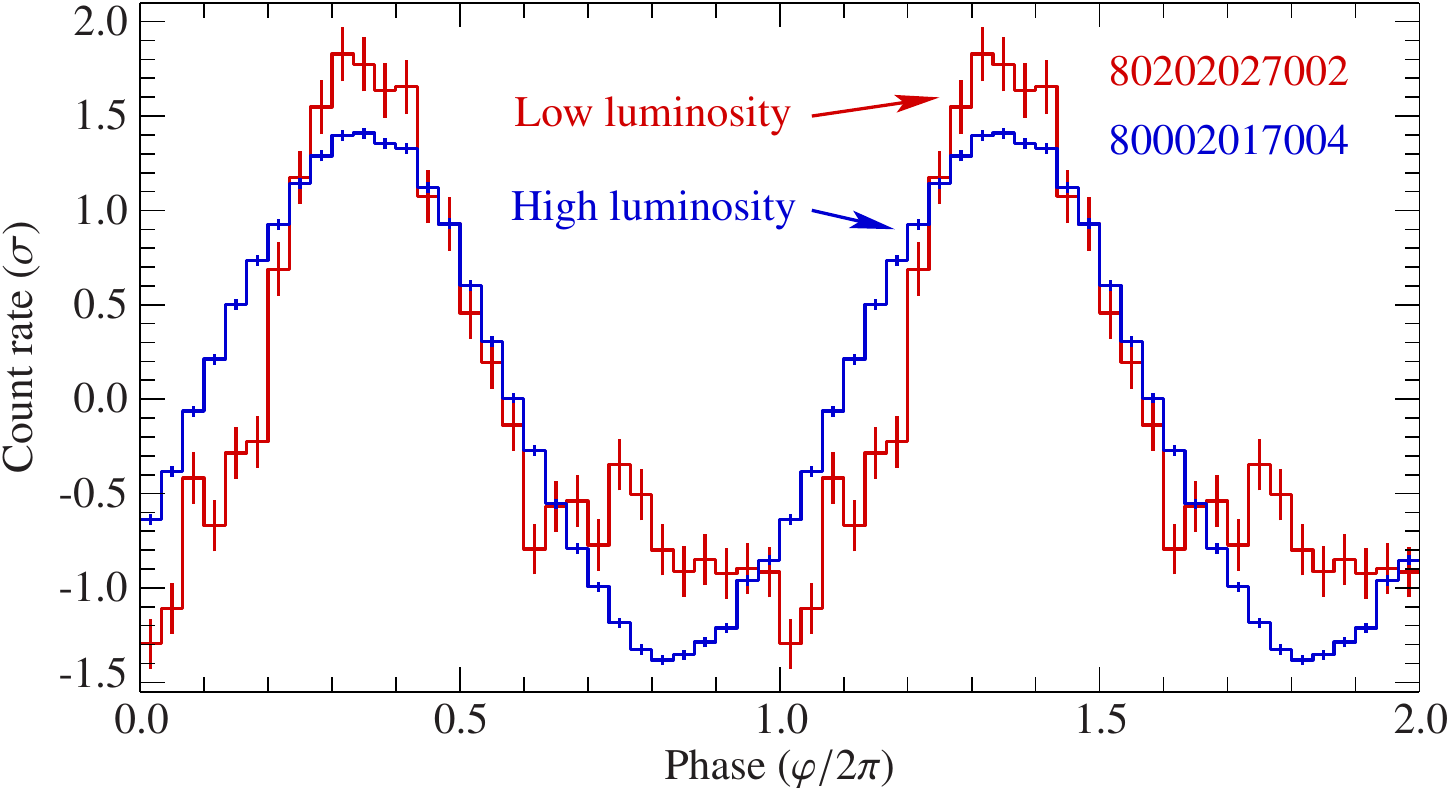}}
  \caption{\nustar pulse profiles of this observation (red) taken at a
    luminosity of $3.2\times 10^{36}$\,\ergs (3--50\,keV) and the 2014 outburst (blue) at
    $1.9\times10^{38}\,\ergs$\,\citep[0.5--10\,\kev,][]{younes15a}.
    The pulse profile of this observation is narrower, less sinusoidal
    and shows the indication of a secondary peak at late phases.
  }
  \label{fig:compPP}
\end{figure}

\section{Discussion \& Summary}
\label{sec:disc}
\subsection{Discussion}
In this paper we presented a spectral analysis of the fourth recorded
outburst of the X-ray pulsar \gro in a low flux state, which occurred
in 2017 February. In contrast to previous outbursts the source only
reached a luminosity of $3.2\times 10^{36}$\,\ergs (3--50\,keV), assuming a distance of 8\,kpc. We
note that this is, however, still three orders of magnitude above
quiescence level \citep{Daigne02a,WijnandsWang02a}.

The spectral shape found during our low luminosity observation can be
well described by an absorbed cut-off power-law with an additional
fluorescence iron emission line complex at $\sim$6.6\,\kev. The
spectral shape is also consistent with a \npex, \fdcut, or
\texttt{nthcomp} model, which were used in some of the earlier studies
of \gro. For this reason, a direct comparison of the model parameters
is difficult. A comparison of the spectral shapes found in earlier
data \citep{doroshenko15a,younes15a,Cui98a}, however shows a slightly
softer spectral shape in this low luminosity observation. This
softening with decreasing luminosity is consistent with previous
studies of \gro \citep{Cui98a,Daigne02a,WijnandsWang02a}, and
generally on accreting pulsars \citep{Reig13a,Postnov15a}. We note
that we do not see a double-humped structure in the spectrum as seen
in a few accreting X-ray pulsars at even lower luminosity
\citep{Tsygankov19b,Tsygankov19a}.

The most debated feature in the X-ray spectrum is the existence of the
cyclotron line. CRSF are difficult to detect below 10\,keV, due to
other spectral features in this regime, and a confirmation of the
$\sim$5\,keV line would make \gro one of the few secure neutron star
systems with a weak magnetic field. The existing claims for CRSFs in
\gro were at a luminosity $2.4\times 10^{38}$\,\ergs (d=8\,kpc),
slightly above its Eddington luminosity \citep{dai15a} and in
super-critical accretion regime \citep{becker12a}. However, even at
these high luminosities, the CRSF was not always seen. Just three days
before the detection of \citet{dai15a}, \citet{younes15a} did not
detect a CRSF, and while \citet[Table~3]{doroshenko15a} observed the
line in the brightest state of the 1997 outburst, it vanished
($\tau=0$) at later times when the luminosity had decreased to
$0.46\times 10^{37}$\,\ergs~(2--10\,keV).

During this \nustar observation \gro had a luminosity of $3.2\times 10^{36}$\,\ergs (3--50\,keV),
two orders of magnitude below the Eddington luminosity. Neither the
phase-averaged nor the phase-resolved spectrum exhibit a significant
cyclotron resonant scattering feature, with a $4\sigma$ upper limit
for the strength of 0.15\,keV. There are several possible reasons for
this vanishing of the cyclotron line. First, in some sources we see
that the line energy depends on luminosity \citep[e.g.,][and references
therein]{2019A&A...622A..61S}. Therefore, it is possible that the line
energy strongly depends on luminosity such that it could be located
below the lower energy threshold of \nustar. Such a dependency would
require a large increase of the height of the accretion column,
however, while we expect accretion columns at low luminosities to be
small \citep[e.g.,][]{becker12a,nishimura:2014,mushtukov:2015}.
Alternatively, the large luminosity change could have changed the
emission geometry such that the conditions at which the line is formed
are not met. Finally, it could also be that the CRSF seen in earlier
data is an artifact of the continuum modeling with simple empirical
models.

Even though the source was at a flux of only $4.15\times 10^{-10}$\,\ergcms, we still detect
pulsations. The pulse profile shows a prominent peak, and a smaller
secondary peak at phase $\sim$0.75 is apparent that was not seen in
the earlier higher luminosity data, which were smooth and almost
perfectly sinusoidal \citep[e.g.,][and
  Fig.~\ref{fig:compPP}]{doroshenko15a,dai15a,younes15a}. The spectral
shape is only very slightly variable with pulse phase.

Our observation of distinct pulsations (and thus the presence of a hot
spot or an accretion column) at such a low flux contradicts earlier
\rxte analyses, where the ``propeller effect'' was claimed to set in
at a flux of $2.34\times 10^{-9}$\,\ergcms \citep{Cui97}. Even though
\rxte had a much larger effective area than \nustar, the background
level in the \rxte PCA was much higher. We therefore speculate that
the non-detection of pulsations at higher flux was due to the lower
signal-to-noise ratio of the earlier observations. With the newer
\nustar data we can therefore revise the threshold for the transition
into the propeller regime to below $4.15\times 10^{-10}$\,\ergcms (3--50\,\kev), i.e., to
almost an order of magnitude below the value found earlier. If we
assume that the resulting luminosity is the transitional threshold for
the propeller regime we can constrain the surface magnetic field
strength to $B\lesssim 2.9\times 10^{11}$\,G (for a canonical neutron star, see
Eq.~\ref{eq:bestimate}) and the mass accretion rate to
$\dot{M}\lesssim 1.7\times 10^{16}$\,g\,s$^{-1}$. The $B$-field estimate is in line with
previous estimates on the source \citep{degenaar14a,younes15a}. We
note, that the value would imply a red-shifted CRSF at
$\lesssim$2.6\,keV, which is outside of \nustar's energy range and
about 2\,keV lower than the claimed CRSF.

We caution that there are several different versions of the propeller
luminosity (Eq.~\ref{eq:bestimate}) in use, depending on the
underlying assumptions about accretion geometry and magnetic field
configuration. For example \citet[Eq.~2 and evaluating their
  $B_0$]{Campana01a} use a pre-factor of $1.69\times 10^{37}$\,\ergs,
\citet[Eq.~4]{Tsygankov17a} of $4\times 10^{37}$\,\ergs, and
\citet[Eq.~2]{Fuerst17a} of $7.3\times 10^{37}$\,\ergs. Furthermore,
different authors use different values of the accretion geometry
parameter $k$. This can result in relatively large differences in the
$B$-field estimate, to be specific $B\propto
(1/\mathrm{prefac})^{1/2}~k^{-7/4}$. If we evaluate the equation by
\citet[$k=1$ and a pre-factor of $4.8\times 10^{37}$\,\ergs]{Cui97} we
obtain a $B$-field of $B\lesssim 1.0\times 10^{11}$\,G, which is a
difference of factor 2.8 to our parameter choice.

Finally, we turn to the occurrence of X-ray bursts. Earlier
observations at persistent fluxes ranging from
$10^{-8}$--$10^{-9}$\,\ergcms
\citep{kouveliotou96a,Jahoda96a,woods99a,younes15a} showed the rate of
X-ray bursts to decrease with flux from $\sim20\,\mathrm{hour}^{-1}$
to $1\,\mathrm{hour}^{-1}$ \citep{kouveliotou96a}. In quiescence, no
X-ray bursts were observed \citep{Daigne02a,WijnandsWang02a}, and
neither did we see evidence for X-ray bursts here. This indicates that
at a flux of $4.15\times 10^{-10}$\,\ergcms, \gro must be in a regime where the burst rate is
less than $\sim 0.1\,\mathrm{h}^{-1}$, possibly even implying that Type~II X-ray
bursts cease below a certain mass accretion rate.

In order to place further constraints on both, the transition flux to
the propeller regime and on the existence of a cyclotron line, further
observations with better \snr-ratio -- for instance with the Neutron
Star Interior Composition Explorer (\nicer) -- are necessary.

\subsection{Summary}
Our most important results of this spectral analysis of the 2017
outburst with \nustar are:
\begin{itemize}
\item \gro had a luminosity of $3.2\times 10^{36}$\,\ergs (3--50\,keV), two orders of
  magnitude below previous outbursts but three orders of magnitude
  above quiescence level.
\item The lightcurve shows no Type~II X-ray bursts which means that the
  burst rate is less than $\sim 0.1\,\mathrm{h}^{-1}$.
\item The powerlaw-shaped spectrum is slightly softer than in the
  high-luminosity case.
\item We cannot find a significant CRSF in the spectrum and put a
  4$\sigma$ upper limit of 0.15\,keV on the \gabs strength.
\item The presence of pulsations allows us to set the threshold for
  the transition into the propeller regime to below $4.15\times 10^{-10}$\,\ergcms, almost an
  order of magnitude lower than previously found. The resulting
  surface magnetic field can be constrained to $\lesssim 2.9\times 10^{11}$\,G
  and the mass accretion rate to $\lesssim 1.7\times 10^{16}$\,g\,s$^{-1}$.
\end{itemize}
  
\begin{acknowledgements}
  OK thanks the ESAC Trainee Program
  (\url{https://www.cosmos.esa.int/web/esac-trainees}) which initiated
  this analysis, and the ERASMUS+ fellowship program for their
  financial support. RB acknowledges funding by Deutsches Zentrum
  f\"ur Luft- und Raumfahrt under contract 50\,OR\,1606. MTW is
  supported by the NuSTAR Guest Investigator Program. This research
  has made use of ISIS functions (ISISscripts) provided by ECAP/Remeis
  observatory and MIT
  (\url{http://www.sternwarte.uni-erlangen.de/isis/}). This research
  has made use of the NuSTAR Data Analysis Software (NuSTARDAS)
  jointly developed by the ASI Science Data Center (ASDC, Italy) and
  the California Institute of Technology (USA).
\end{acknowledgements}

\bibliographystyle{jwaabib}
\bibliography{mnemonic,aa_abbrv,bibfiles}

\end{document}